\documentclass[final,5p,times,twocolumn]{elsarticle}
\usepackage[mathscr]{euscript}
\usepackage{dblfloatfix}
\usepackage{bm}
\usepackage{textcomp}
\usepackage{graphicx}
\usepackage{subfigure}
\usepackage{overpic}
\usepackage{multirow}
\usepackage{floatrow}
\usepackage{float} 
\usepackage{amsmath,amssymb,amsthm}
\usepackage{cases}
\usepackage{cancel}
\usepackage[colorlinks=true,linkcolor=red]{hyperref}
\usepackage{xcolor}
\biboptions{sort&compress}
\journal{Physics Letters B}
\newfloatcommand{capbtabbox}{table}[][\FBwidth]
\newcommand{\bea}{\begin{eqnarray}}
\newcommand{\eea}{\end{eqnarray}}
\newcommand{\beq}{\begin{equation}}
\newcommand{\eeq}{\end{equation}}

\def\/{\over}

\begin{document}

\title{Hawking evaporation and the possibility of remnant formation for
covariant quantum black holes
}

\author
{Li-Shuai Wang}
\author{Xiangdong Zhang\corref{cor1}}
\cortext[cor1]{Corresponding author: scxdzhang@scut.edu.cn}

\address{
  School of Physics and Optoelectronics, South China University of Technology, Guangzhou 510641, China
}

\begin{abstract}

Covariant quantum black holes (BHs) are an important development in loop quantum gravity (LQG), and investigations of their evaporation may offer a possible avenue for testing LQG. In this paper, we investigate the evaporation of covariant quantum BHs and the possibility of remnant formation. Our results show that covariant quantum BHs can leave remnants, calculated by analytical greybody factors for a massless scalar field. Nevertheless, the fully numerical results indicate  mass loss rates of covariant quantum BHs are spin-dependent and differ from those of Schwarzschild BHs, due to spin-dependent greybody factors. In particular, these BHs evaporate faster than Schwarzschild BHs in the low-mass regime, with the evaporation being dominated by massless neutrino emission rather than scalar field emission, contrary to the scalar field dominance commonly assumed in previous studies. Therefore, considering only BH evaporation in the massless scalar field case may be insufficient and may even lead to misjudging BH fate. In addition, these results may provide a way to test LQG in the future.

\end{abstract}



\maketitle
\section{Introduction}
\label{sec_in}
General relativity has achieved remarkable success in explaining many problems and phenomena. In modern cosmology, it provides the theoretical foundation for the $\Lambda$CDM model, which can account for the accelerated expansion of the universe and is consistent with observations of the cosmic microwave background~\cite{Planck:2018vyg}. In the strong field regime, it predicted the existence of black holes (BHs) which  were evidenced by the BH image released in 2019~\cite{EventHorizonTelescope:2019dse}, and gravitational waves which were observed by LIGO~\cite{LIGOScientific:2016aoc}.  However, general relativity cannot explain several key problems, such as the BH information paradox and spacetime singularities~\cite{PhysRevLett.14.57,PhysRevD.14.2460}. A number of physical phenomena and theoretical requirements indicate quantum gravity theory is necessary. Nevertheless, due to unresolved conceptual and technical difficulties, a complete and universally accepted quantum gravity theory is still absent. 

Even so, several important candidate theories are being developed in this direction. One prominent candidate is string theory, which considers gravity as a fundamental interaction and regards elementary particles as one-dimensional vibrating strings~\cite{Schwarz:2007yc}. Different types of strings and their distinct vibrational modes can account for different elementary particles. String theory has achieved significant success for explaining a variety of phenomena and theoretical problems~\cite{Alvarez-Gaume:2022aak,Chen:2021lnq}. Another highlighted candidate is loop quantum gravity (LQG), which more closely follows the geometric spirit of general relativity by attempting to quantize spacetime geometry itself in a background independent way ~\cite{Ashtekar:1986yd,Rovelli:1989za,Rovelli:1994ge}. LQG has made notable progress in several directions, such as loop quantum cosmology~\cite{Ashtekar:2006rx,Yang:2009fp,Assanioussi:2018hee,Shuai:2026akn}, and quantum-corrected BH models~\cite{Peltola:2008pa,Bianchi:2018mml,DAmbrosio:2018wgv,Lewandowski:2022zce}. In addition, considerable progress has been made in the studies of quantum Oppenheimer-Snyder BHs and covariant quantum BHs in recent years~\cite{Lewandowski:2022zce,Shi:2024vki,Chen:2025baz,Belfaqih:2024vfk,Du:2025kcx,Ou:2025bbv,Lin:2024beb,Du:2024ujg,Tan:2025hxc}. 

For quantum-corrected
BHs, quantum corrections modify the BH spacetime geometry, causing it to deviate from the classical spacetime, such as the Schwarzschild spacetime. It may lead to the modifications to BH temperatures and greybody factors, thereby changing the intensity of Hawking radiation and the BH mass loss rate. Relatedly, quantum Oppenheimer-Snyder BHs and a class of covariant quantum BHs have been considered as primordial BHs~\cite{Wang:2026axy,Calza:2025mwn}, which have different observational constraints compared with Schwarzschild BHs, due to the quantum corrections to the Hawking evaporation. Therefore, to investigate the BH evaporation may be a key way for exploring quantum corrected BHs and testing quantum gravity theories. In addition, another class of covariant quantum BHs may leave remnants in the massless scalar field case because of quantum corrections to the evaporation process~\cite{Belfaqih:2025eak}. However, real‑world BHs emit not only scalar particles but also spin-1/2 fermions (neutrinos). For semi‑classical Hawking evaporation, massless neutrino emission, with
all degrees of freedom taken into account, is not weaker than massless scalar field emission for Schwarzschild BHs~\cite{Arbey:2021jif}. It is therefore essential to check whether the remnant phenomenon survives for covariant quantum BHs when these spin‑1/2 degrees of freedom are incorporated. This motivates us to investigate the evaporation and fate of covariant quantum BHs by taking into account particles of all relevant spins. 

Based on these considerations, we investigate the evaporation and fate of the class of covariant quantum BHs~\cite{Zhang:2024khj}, whose metric is time-radial symmetric. The spacetime metric contains additional quantum correction terms compared with the Schwarzschild metric. When the quantum correction parameter is close to zero, the spacetime reduces to the Schwarzschild spacetime. These quantum correction terms cause the spacetime geometry to deviate from the Schwarzschild spacetime. This deviation may further lead to modifications to the greybody factors and the BH temperature, and consequently result in Hawking radiation and mass loss rates different from those of Schwarzschild BHs. 

Our results show that the temperature of the covariant quantum BHs is consistent with the Schwarzschild temperature, and greybody factors are almost the same as those of Schwarzschild BHs in the large mass regime, leading to the Hawking radiation and the mass loss rates that are nearly consistent with Schwarzschild results. But in the low-mass regime, mass loss rates of the covariant quantum BHs can be larger than, smaller than, or comparable to classical limits, depending on the spin of
the emitted particles. In particular, the covariant quantum BHs may leave remnants when only massless scalar field emission is considered. However, the effects of quantum corrections in the massless neutrino case are opposite to those in the massless scalar field case, and the quantum-corrected evaporation is even dominated by massless neutrino emission, with the total evaporation proceeding faster than classical evaporation. These results indicate considering only the contribution from a single spin case may be insufficient and may even lead to an incorrect conclusion. In addition, since the Hawking radiation and the mass loss rates of
the covariant quantum BHs are different from those of Schwarzschild BHs, these differences
may provide a potential observational window for testing LQG in the future. 

This paper is organized as follows: Sec.~\ref{sec:Covariant quantum BH} briefly reviews the class of covariant quantum BHs and calculates their temperature. Sec.~\ref{sec:greybody factors} takes the massless scalar field and massless neutrino cases as examples to numerically calculate and compare the  corresponding greybody factors. Sec.~\ref{sec:Hawking radiation} numerically investigates the Hawking radiation of the covariant quantum BHs and Schwarzschild BHs in the massless scalar field and the massless neutrino cases. Sec.~\ref{sec:Mass loss rate} computes and compares the mass loss rates of Schwarzschild and the covariant quantum BHs and discusses the possibility of remnant formation. In the end, Sec.~\ref{sec:conclusions} summarizes the conclusions of this paper. Moreover, we set $c=G=\hbar=k_{B}=1$ in this paper.

\section{Covariant quantum BH}
\label{sec:Covariant quantum BH}
In this section, we simply review the class of covariant quantum BHs proposed in Ref.~\cite{Zhang:2024khj}, whose spacetime metric is time-radial symmetric. For a static, spherically symmetric spacetime, the line element, in Boyer-Lindquist coordinates \((t, r, \theta, \phi)\), is as follows
\begin{equation}
ds^2 = -f(r)dt^2 + g(r)^{-1}dr^2 + h(r)d\Omega^2\,,
\label{eq:ds2}
\end{equation}
where
\begin{equation}\label{f(r)}
f(r)=1 - \frac{2M}{r} + \frac{\zeta^2}{r^2} \left(1 - \frac{2M}{r} \right)^2,
\end{equation}
\begin{equation}\label{g(r)}
g(r)=1 - \frac{2M}{r} + \frac{\zeta^2}{r^2} \left(1 - \frac{2M}{r} \right)^2,
\end{equation}
and
\begin{equation}\label{h(r)}
h(r)=r^2.
\end{equation}
Here, $M$ is the BH mass, $d\Omega^2=d\theta^2+\sin^2\theta d\phi^2$, the quantum parameter is defined as $\zeta = \sqrt{4\sqrt{3}\pi \gamma^3 \ell_p^2}$, with \( \gamma \) being the Barbero-Immirzi parameter, and \( \ell_p \) is the Planck length. When $\zeta$ is close to zero, the quantum corrected spacetime reduces to Schwarzschild spacetime. The quantum corrections modify the spacetime geometry, so the temperature of the covariant quantum BHs may be different 
from that of Schwarzschild BHs. In addition, since the BH temperature is a key element of Hawking evaporation, it is necessary to obtain it for calculating Hawking evaporation. 

The Hawking temperature of a static, spherically symmetric BH is defined by the surface gravity which is calculated at the horizon. In terms of the surface gravity $\kappa$, the BH temperature is given by
\begin{equation}\label{surface gravity}
T=\frac{\kappa}{2\pi}.
\end{equation}
The surface gravity can be computed from the near horizon metric. For a static, spherically symmetric BH, the surface gravity is calculated by
\begin{equation}\label{kappa}
\kappa=\sqrt{\frac{g(r)}{f(r)}}\frac{f'(r)}{2}\vert_{r_H}.
\end{equation}
Here, $r_{H}$ is the horizon radius. Therefore, in the spacetime considered here, the BH temperature can be calculated by 
\begin{equation}\label{temperature}
T=\sqrt{\frac{g(r)}{f(r)}}\frac{f'(r)}{4\pi}\vert_{r_H}\,.
\end{equation}
It is easy to see that the horizon radius plays a key role in evaluating the BH temperature. Before computing the BH temperature, it is necessary to obtain the horizon radius. For the spacetime metrics considered here, the horizon radius $r_H=2M$ can be obtained by imposing $f(r_H)=g(r_H)=0$. Combining Eqs.~(\ref{f(r)}),~(\ref{g(r)}), and~(\ref{temperature}), and using $r_H=2M$, one can obtain $T=1/(8\pi M)$. It is obvious that the temperature of the class of covariant quantum BHs is the same as that of Schwarzschild BHs. This result shows that quantum corrections do not modify the temperature of the covariant quantum BHs compared with Schwarzschild BHs. Nevertheless, the quantum corrections may still modify the greybody factors, leading to modifications of the Hawking evaporation. Based on this expectation, possible modifications of the greybody factors will be investigated in the next section.

\section{greybody factors}
\label{sec:greybody factors}
BHs can induce effective potentials around them by gravity, so the particles which are emitted from near the BH horizon may penetrate the effective potential and can finally arrive at infinity. Therefore, the Hawking radiation is greybody radiation rather than blackbody radiation observed at infinity. The transmission coefficients of particles passing through the effective potential are usually referred to as greybody factors. From the previous section, we have found that the quantum corrections do not modify the temperature of the class of covariant quantum BHs. However, the quantum corrections can modify spacetime geometry, which may affect greybody factors. Since greybody factor is another important element for Hawking evaporation, it needs to be obtained. In addition, the radial Teukolsky equation should be solved for calculating greybody factors and the Teukolsky equation for the radial component of a field with spin $s$, $R_{s}$, in a static, spherically symmetric spacetime, can be written as~\cite{Arbey:2021jif,Calza:2024xdh}
\begin{equation}\label{Teukolsky1} 
A_s(B_s R'_s)'+ \left [ \frac{h}{f}\omega^2+i\omega s\sqrt{\frac{g}{f}} \left ( h'-\frac{h f'}{f} \right ) + C_s \right ] R_s=0\,, \\
\end{equation}
where
\begin{equation}
A_s= \sqrt{\frac{g}{f}} \frac{1}{(f h)^s}\,,
\label{eq:as}
\end{equation}
\begin{equation}
B_s=\sqrt{f g } (f h)^sh\,,
\end{equation}
\begin{align}
C_s &= s \frac{g h f''}{f} + \frac{s}{2} \left ( \frac{h g' f' }{f} - \frac{g h f'^2}{f^2} \right ) \nonumber \\ 
&+ \frac{s(3-2s)}{4} \left( 2 g h'' + g' h'  \right) +\frac{s(2s-1)}{4} \frac{g h'^2}{h}\nonumber \\ 
&+\frac{s(2s+5)}{4}\frac{g f' h'}{f}-\nu^s_l-2s\,.
\end{align}
Here, $\omega$ is the frequency of particles, a prime represents a derivative with respect to $r$, $\nu^s_l=l(l+1)-s(s-1)$, $s$ is the spin of particles, and $l$ is the angular momentum quantum number. Greybody factors can be calculated by directly solving the radial Teukolsky equation. This method is usually called the shooting method and more details can be found in Refs.~\cite{Calza:2024xdh,Calza:2024fzo}. However, to show more clearly how the effective potential affects greybody factors, we will adopt an approach in which the radial Teukolsky equation is rewritten as a Schr\"odinger-like equation with effective potentials to calculate greybody factors. After suitable transformations, the radial Teukolsky equation can be written as follows~\cite{Arbey:2021yke,Arbey:2021jif}
\begin{equation} \label{eq:SchEq}
    \partial^2_{r_*}Z_s + \Bigl(\omega^2-V_{sl} \Bigr)Z_s=0,
\end{equation}
where $Z_s$ is the wave function, and $V_{sl}$ are effective potentials, for several spin cases, as follows
\begin{subequations}\label{eq:potentials}
\begin{align}
	&V_{0l}=\nu^0_l\frac{f}{h}+\frac{\partial_{r_*}^2\sqrt{h}}{\sqrt{h}}\,,\\
	&V_{1l}=\nu^1_l\frac{f}{h}\,,\\
	&V_{2l}=\nu^2_l\frac{f}{h}+\frac{(\partial_{r_*}h)^2}{2h^2}-\frac{\partial_{r_*}^2\sqrt{h}}{\sqrt{h}}\,,\\
	&V_{\frac12l}=\nu^{\frac12}_l\frac{f}{h} \pm \sqrt{\nu^{\frac12}_l}\,\partial_{r_*}\left( \sqrt{\frac{f}{h}} \right)\,.
\end{align}\label{pot}
\end{subequations}

\noindent To further evaluate the wave function $Z_s$, we need to define the tortoise coordinate $r_*$ as follows
\begin{equation}\label{eq:tort}
    \frac{dr_*}{dr} = \frac{1}{\sqrt{f(r) g(r)}}.
\end{equation}

The effective potentials vanish near the horizon and at infinity, so Eq.(\ref{eq:SchEq}) has plane wave solutions. Since we assume particles originate from infinity and propagate toward the BH, ingoing waves and outgoing waves are present at infinity. In a similar way, only ingoing waves are admitted near the horizon. Thus, near the horizon, the wave function $Z_s$ is given by
\begin{equation}\label{near-horizon}
    Z_s(r_*) \underset{r_*\rightarrow -\infty}{\sim} A_{hor}^{\rm in} e^{-i\omega r_*}\,,
\end{equation}
and, at infinity, $Z_s$ can be described as follows
\begin{equation}\label{Zfar}
    Z_s(r_*) \underset{r_*\rightarrow +\infty}{\sim} A_{\infty}^{\rm in}e^{-i\omega r_*} + A_{\infty}^{\rm out}e^{+i\omega r_*}\,,
\end{equation}
Here, $A_{hor}^{\rm in}$, $ A_{\infty}^{\rm in}$, and $A_{\infty}^{\rm out}$ are constants. 

\begin{figure}[!h]
\centering
\begin{minipage}{0.9\linewidth}
\centering
\includegraphics[width=\linewidth]{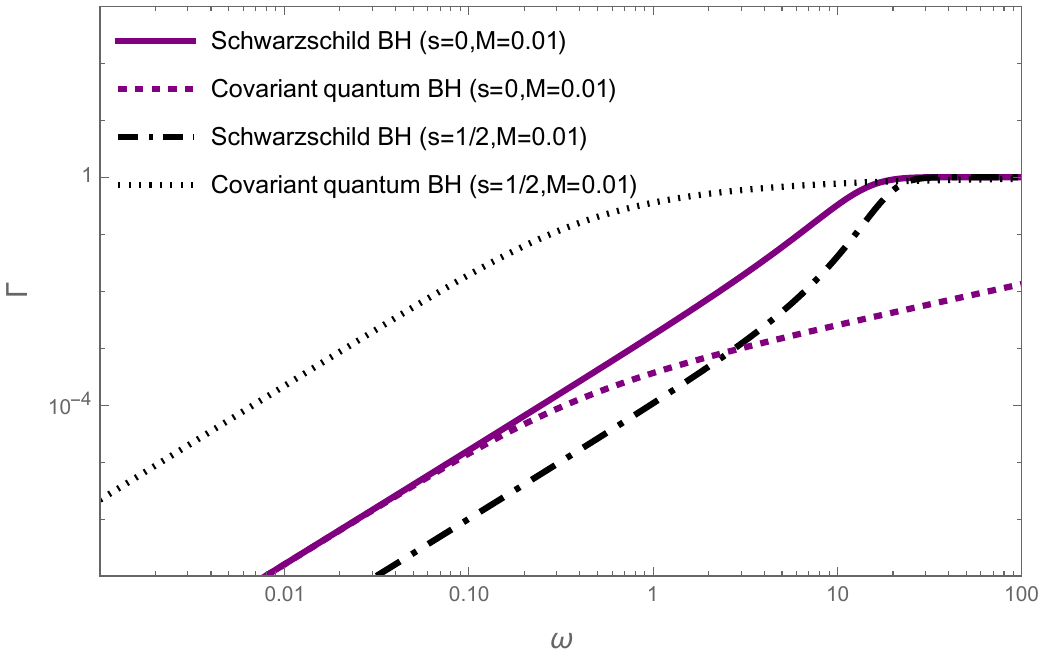}\\[-0.25cm]
(a)
\end{minipage}
\\[0.3cm]
\begin{minipage}{0.9\linewidth}
\centering
\includegraphics[width=\linewidth]{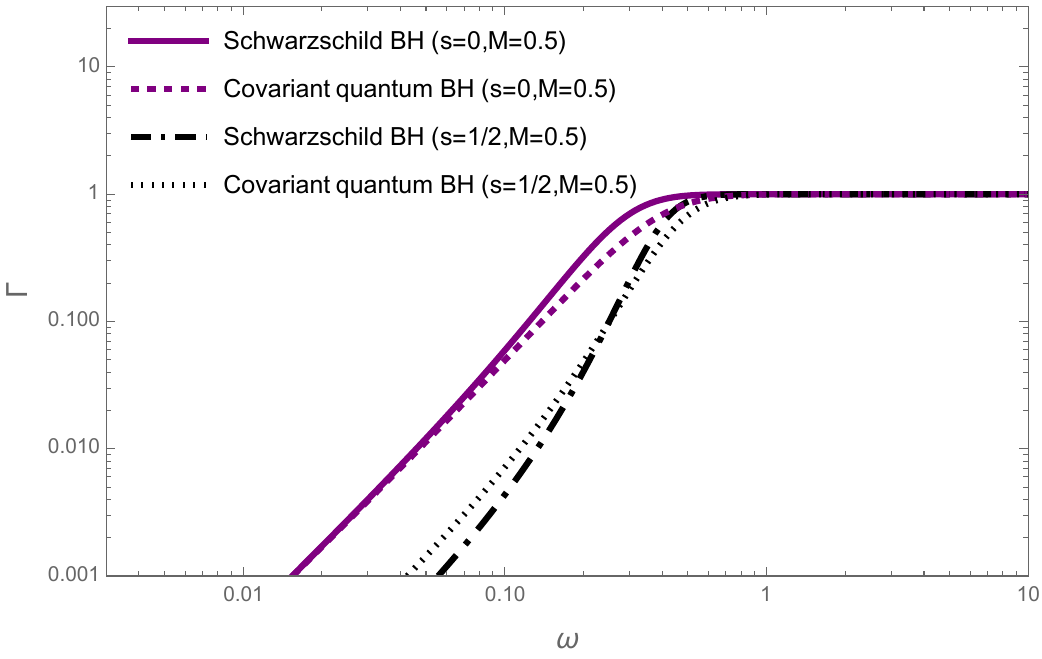}\\[-0.25cm]
(b)
\end{minipage}
\caption{Greybody factors as functions of $\omega$. For $s=0$, the solid purple line represents the greybody factors for Schwarzschild BHs, while the dashed purple line represents the greybody factors for the covariant quantum BHs. For $s=1/2$, the black dot-dashed line and black dotted line are the greybody factors for Schwarzschild BHs and covariant quantum BHs, respectively. In addition, the greybody factors for BH masses $M=0.01$ and $M=0.5$ are shown in (\textbf{a}) and (\textbf{b}), respectively. We set $l=0$ for $s=0$ and $l=1/2$ for $s=1/2$, with $\zeta=\sqrt{3}$ in this figure.}
\label{fig1}
\end{figure}

\noindent Since the greybody factors are transmission coefficients here, they can be evaluated by
\begin{equation}\label{greybody factors}
    \Gamma^i_{lm}(\omega,M,x_j) = \left|\dfrac{A_{hor}^{\rm in}}{A_{\infty}^{\rm in}}\right|^2\,,
\end{equation}
where the index $i$ labels the particle species, $x_j$, denoted by $\zeta$ in this paper, is the set of precise shape and parameters of the metric, and $m$ is the magnetic quantum number which is degenerate for a given $l$ in a spherically symmetric spacetime.

We numerically calculate and plot the greybody factors for the massless scalar field and massless neutrino cases in Fig.~\ref{fig1}(a) and Fig.~\ref{fig1}(b) with $M=0.01$ and $M=0.5$, respectively. It is evident that when $M=0.5$, the greybody factors for the covariant quantum BH are close to the Schwarzschild results. However, when $M=0.01$, a number of greybody factors for the covariant quantum BH obviously deviate from their classical limits. The greybody factors for a massless scalar field are smaller than the Schwarzschild results at certain frequencies. But for massless neutrino emission, the greybody factors are larger than the Schwarzschild results and quantum correction results for a massless scalar field at certain frequencies.  These results show that the quantum corrections become stronger for smaller BH masses and may lead to opposite effects for different spin cases.

\section{Hawking radiation}
\label{sec:Hawking radiation}

Particles are emitted through thermal radiation from BHs, but only a fraction can penetrate the potential barrier and arrive at infinity. Thus, the Hawking radiation is primarily determined by BH temperature and greybody factors. In previous sections, it has been known that the temperature of the class of covariant quantum BHs coincides with that of Schwarzschild BHs and the quantum corrections markedly change spacetime geometry, leading to the modifications of the greybody factors for the massless scalar field and massless neutrino cases in the low-mass regime. Therefore, it can be reasonably inferred that the Hawking radiation can still be modified by substantial changes in the greybody factors in the low BH mass regime, even in the absence of any modifications to the temperature. Based on these analyses, it is necessary to evaluate the Hawking radiation, and the number of particles emitted by the Hawking radiation (per unit time and per unit energy) can be calculated by~\cite{Hawking:1975vcx,Page:1976df} 
\begin{equation}\label{prim}
\frac{d^2N_i}{dtdE}=\frac{1}{2\pi}\sum_{l,m}\frac{n_i \Gamma^i_{lm}(\omega,M,x_j)}{ e^{\omega/T}\pm 1}\,,
\end{equation}
where $N_i$ is the number of emitted particles of species $i$, $t$ is the asymptotic coordinate time, and $n_i$ denotes the number of internal degrees of freedom of that species. For massless scalar particles, massless neutrinos, photons, and gravitons, the corresponding internal degrees of freedom are taken to be \(n_i=1,6,2,\) and \(2\), respectively~\cite{Arbey:2019mbc}. To further evaluate the Hawking radiation, $l$ and $m$ need to be summed. However, the magnetic quantum number $m$ is degenerate in spherically symmetric spacetimes, so the Hawking radiation Eq.~(\ref{prim}) can be written as
\begin{equation}\label{prim2}
\frac{d^2N_i}{dtdE}=\frac{1}{2\pi}\sum_{l}(2l+1)\frac{n_i\Gamma^i_{l}(\omega,M,x_j)}{ e^{\omega/T}\pm 1}\,.
\end{equation}

\begin{figure}[h]
\centering
\begin{minipage}{0.9\linewidth}
\centering
\includegraphics[width=\linewidth]{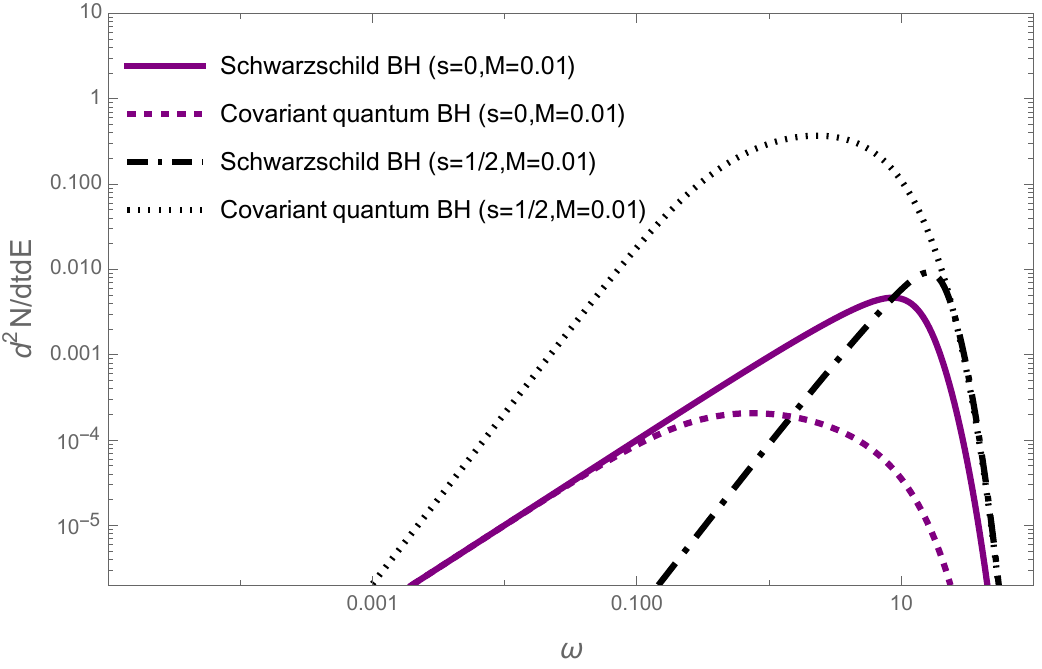}\\[-0.25cm]
(a)
\end{minipage}
\\[0.3cm]
\begin{minipage}{0.9\linewidth}
\centering
\includegraphics[width=\linewidth]{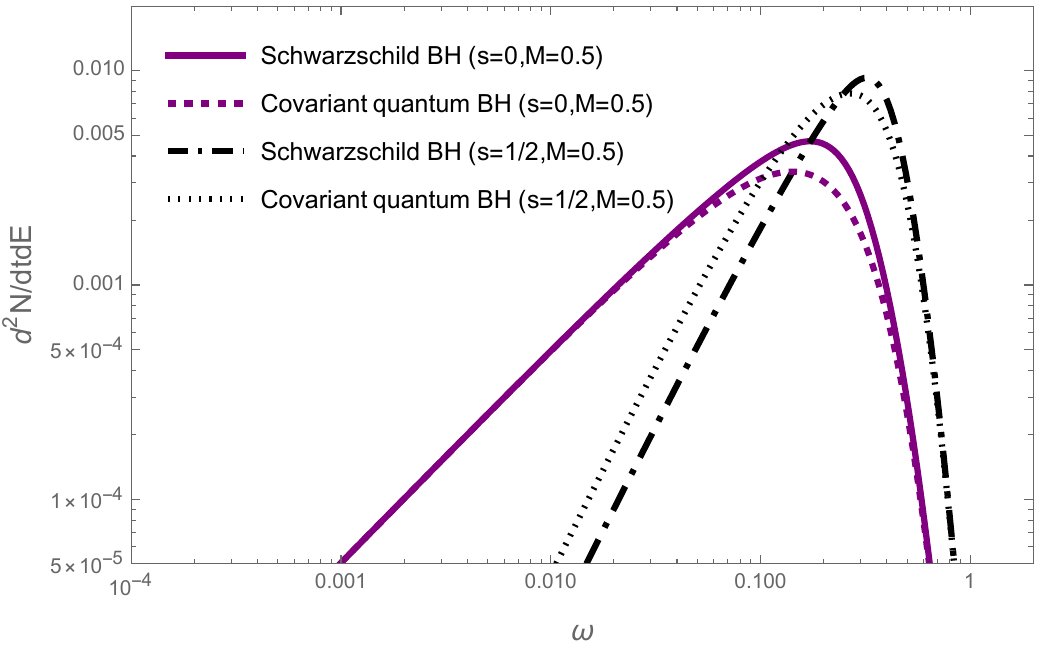}\\[-0.25cm]
(b)
\end{minipage}
\caption{The particle number spectra of the Hawking radiation as functions of particle frequency $\omega$. For $s=0$, the purple solid line is the spectrum of the Schwarzschild BH, and the purple dashed line is the spectrum of the covariant quantum BH. For $s=1/2$, the black dotted line and black dot-dashed line represent the spectrum of the covariant quantum BH and Schwarzschild BH, respectively. (\textbf{a}) The spectrum for BH mass $M=0.01$. (\textbf{b}) The spectrum for BH mass $M=0.5$. We set $l=0$ for $s=0$ and $l=1/2$ for $s=1/2$, with $\zeta=\sqrt{3}$ in this figure.}
\label{fig2}
\end{figure}

Following the discussion of the greybody factors in the previous section, we still focus on $s=0$ and $s=1/2$ as examples and numerically compute the Hawking radiation. Using Eq.~(\ref{prim2}), we plot the particle number spectra of the Hawking radiation for the Schwarzschild BH and the covariant quantum BH with $M=0.01$ and $M=0.5$, as shown in Fig.~\ref{fig2}(a) and Fig.~\ref{fig2}(b), respectively. As can be clearly seen, for BH mass $M=0.01$, the Hawking radiation spectra of the covariant quantum BH significantly deviate from those of the Schwarzschild BH in a certain frequency range, while the peak of the spectrum of the covariant quantum BH for massless neutrino emission is higher than those of the other spectra in Fig.~\ref{fig2}(a). In contrast, for BH mass $M=0.5$, the Hawking radiation spectra of the covariant quantum BH are comparable to the corresponding spectra of the Schwarzschild BH. These results are also consistent with our previous expectation. In the small-mass regime, the quantum corrections significantly affect the greybody factors, thereby leading to substantial modifications of the Hawking radiation intensity. In contrast, in the large-mass regime, the influence of the quantum corrections on the greybody factors becomes negligible, so that the Hawking radiation remains almost unchanged.

\section{Mass loss rate}
\label{sec:Mass loss rate}

Based on the discussions above, the quantum corrected Hawking radiation spectra deviate significantly from their classical counterparts at certain frequencies in the low-mass regime for $s=0$ and $s=1/2$, due to significant modifications of some greybody factors by quantum corrections. When the particle emitted from a BH arrives at infinity, the emitted particle carries away energy from the BH, leading to the loss of BH mass. Therefore, it is natural to infer that the mass loss rate will be obviously modified. If the particle number emitted per unit time and per unit energy is obtained at infinity, multiplying it by the particle energy gives the energy flux spectrum. In this way, the BH mass loss rate will be obtained by integrating over the energy from zero to infinity, and it can be described as follows~\cite{Arbey:2019mbc,MacGibbon:1991tj}
\begin{equation}\label{dM/dt1}
\frac{dM}{dt}=-\int^{+\infty}_0E\frac{d^2N_i}{dtdE}dE,
\end{equation}

\noindent where the minus sign denotes the BH mass is reduced. Combining Eqs.~(\ref{prim2}) and~(\ref{dM/dt1}), the final mass loss rate equation, for a static, spherically symmetric BH, can be given by
\begin{equation}\label{dM/dt2}
\frac{dM}{dt}=-\int^{+\infty}_0\frac{E}{2\pi}\sum_{l}(2l+1)\frac{n_i\Gamma^i_{l}(\omega,M,x_j)}{ e^{\omega/T}\pm 1}dE,
\end{equation}

\subsection{Analytical approximation and remnant formation}

\begin{figure}[h]
\centering
\includegraphics[width=1\linewidth]{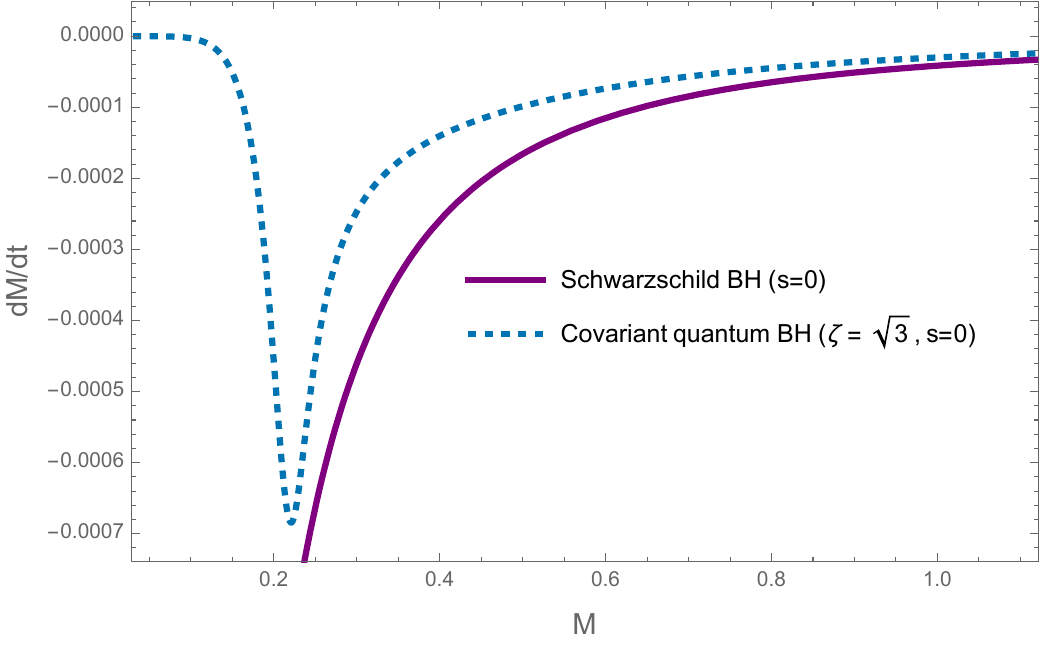}
\caption{The mass loss rate calculated using analytical greybody factors as a function of BH mass for a massless scalar field. The solid purple line and the dashed blue line represent the classical and quantum cases, respectively.}\label{fig3}
\end{figure}

In this subsection, we discuss the possibility of the remnant formation for the covariant quantum BH. In the low-frequency approximation, the analytical greybody factors for a massless scalar field can be derived as follows (see~\ref{Appendix A} for more details)
\begin{equation}\label{analitical greybody factors}
\Gamma=\frac{16M^2\omega^2}{M^2\omega^2(2+\frac{\zeta^2}{3M^2}-\frac{\zeta^4}{160M^4})^2+(1+4M^2\omega^2)^2}.
\end{equation}
Combining Eqs.~(\ref{dM/dt2}),~(\ref{analitical greybody factors}), and~(\ref{Sch greybody factors}), we plot the mass loss rates of the covariant quantum BH and the Schwarzschild BH in Fig.~\ref{fig3}.

One can find mass loss rates for the two types of BHs are nearly identical in the large-mass regime. However, in the low-mass regime, the quantum result deviates from its Schwarzschild counterpart. In particular, when BH mass decreases below approximately $M=0.2$, the mass loss rate of covariant quantum BHs bounces and gradually approaches zero. This result shows that the quantum corrections significantly change the mass loss rates of BHs and the covariant quantum BHs may leave remnants.

\subsection{Fully numerical results for mass loss rates}

Although the semi-analytical result indicates the covariant quantum BHs may leave remnants, the effects of quantum corrections in the massless scalar field and massless neutrino cases are opposite, as shown by the fully numerical results in Figs.~\ref{fig1}(a) and~\ref{fig2}(a). This indicates that it is not sufficiently rigorous to conclude that such covariant quantum BHs will leave remnants without taking massless neutrino emission into account. Therefore, it is necessary to investigate the massless neutrino emission as well as other spin fields. In this subsection, we therefore discuss BH mass loss rates by fully numerical calculations.

\begin{figure*}[!tb]
\centering
\begin{minipage}{0.49\linewidth}
\centering
\includegraphics[width=\linewidth]{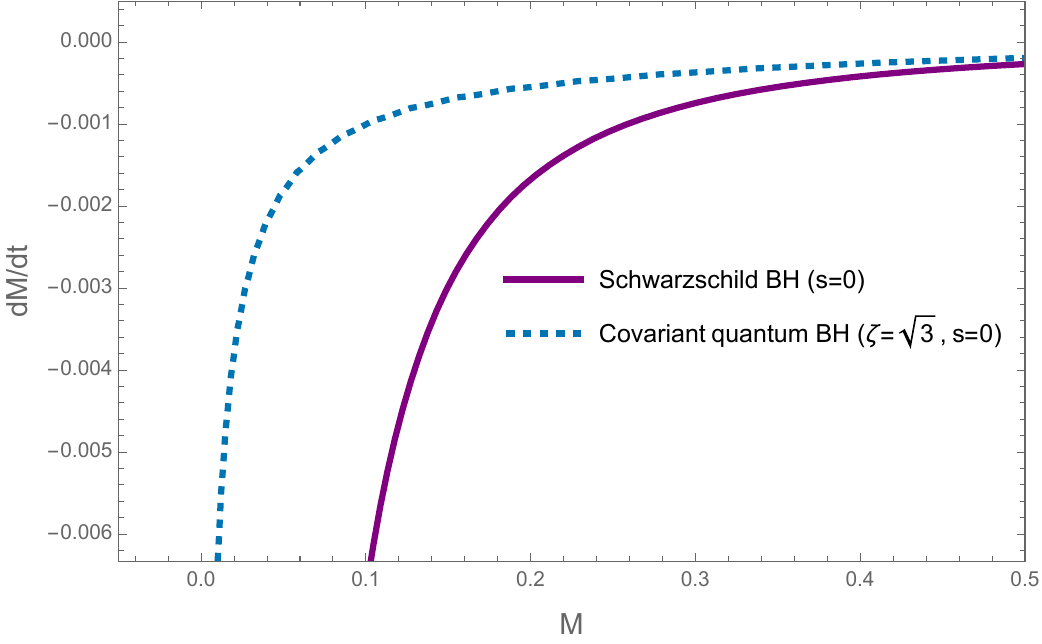}\\[-0.25cm]
(a)
\end{minipage}
\begin{minipage}{0.49\linewidth}
\centering
\includegraphics[width=\linewidth]{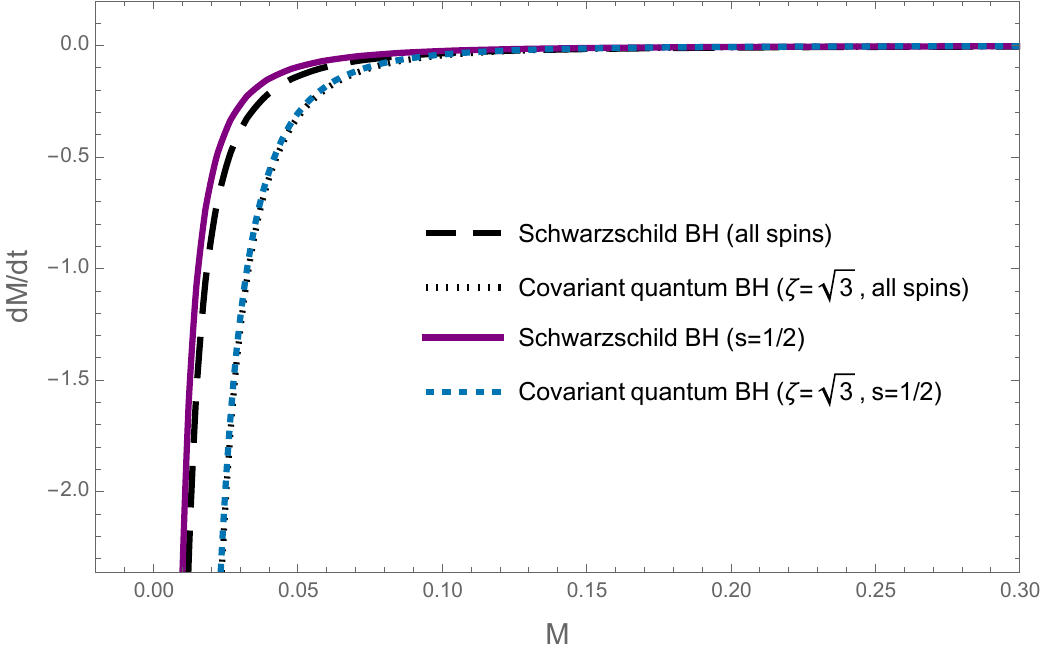}\\[-0.25cm]
(b)
\end{minipage}
\\[0.5cm]
\begin{minipage}{0.49\linewidth}
\centering
\includegraphics[width=\linewidth]{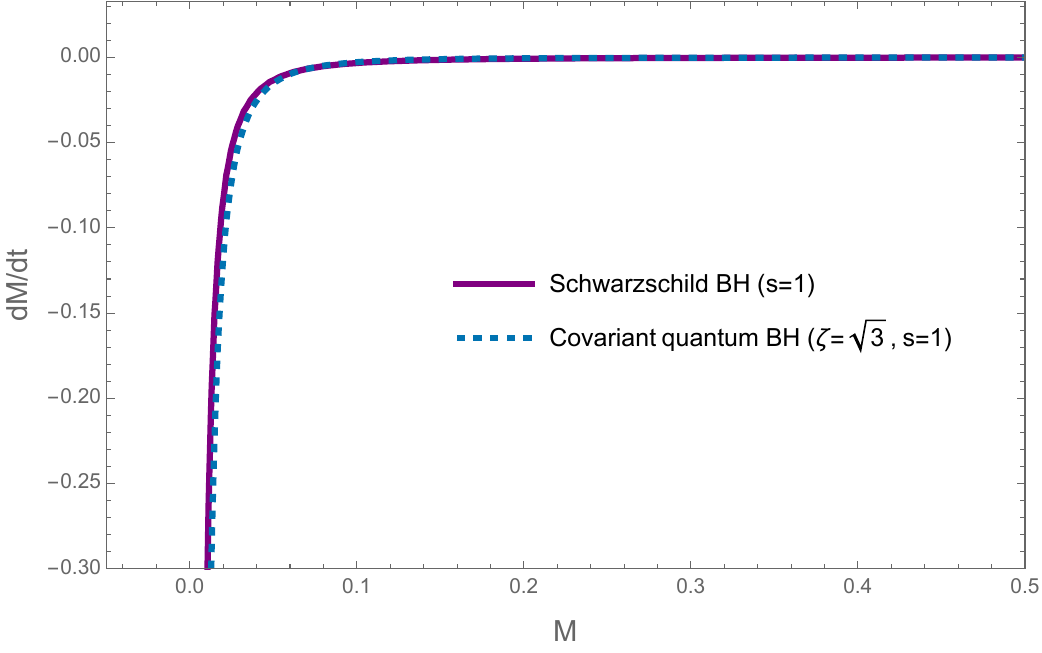}\\[-0.25cm]
(c)
\end{minipage}
\begin{minipage}{0.49\linewidth}
\centering
\includegraphics[width=\linewidth]{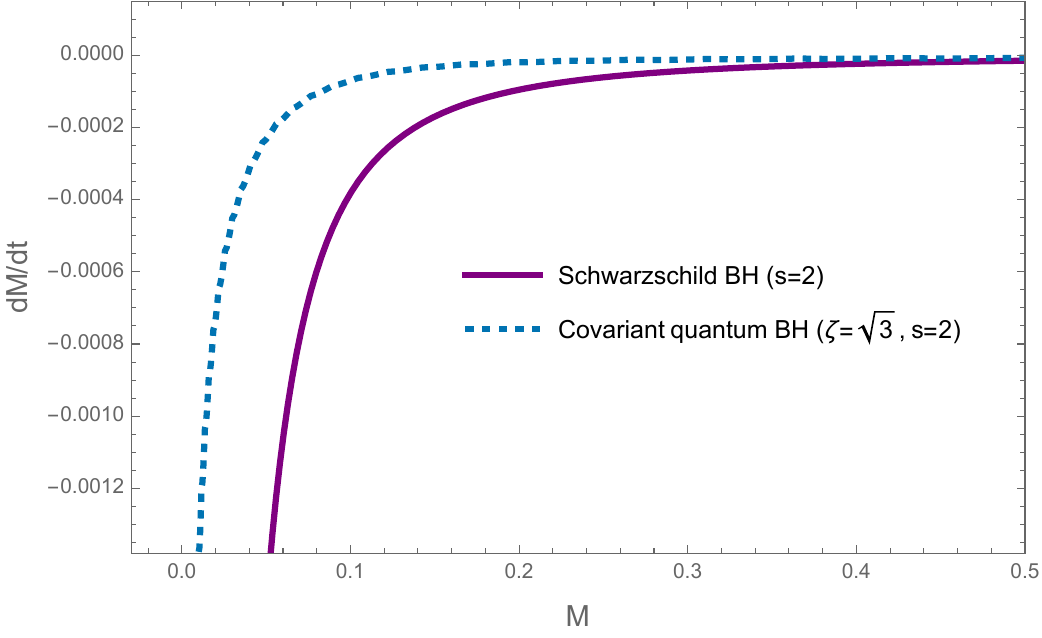}\\[-0.25cm]
(d)
\end{minipage}
\caption{The mass loss rates as a function of BH mass. The purple solid line and the blue dashed line denote the mass loss rates of Schwarzschild BHs and the covariant quantum BHs for the corresponding particle spin, respectively. In addition, the mass loss rates for the massless scalar field, massless neutrino, photon, and graviton cases with $l=0$, $l=1/2$, $l=1$, $l=2$ are shown in panels (\textbf{a}), (\textbf{b}), (\textbf{c}), and (\textbf{d}), respectively. In panel (\textbf{b}), the total mass loss rates of Schwarzschild BHs and covariant quantum BHs are shown by the black long-dashed line and black dotted line, respectively.}
\label{fig4}
\end{figure*}

Using Eq.~(\ref{dM/dt2}), we plot mass loss rates of the covariant quantum BH and the Schwarzschild BH in Fig.~\ref{fig4}(a) for the massless scalar field case. It is obvious that the magnitudes of the mass loss rates of the covariant quantum BHs are significantly smaller than those of Schwarzschild BHs in the low-mass regime. In the same way, we calculate the mass loss rates of the covariant quantum BHs and Schwarzschild BHs for massless neutrino, photon, and graviton cases in Fig.~\ref{fig4}(b), Fig.~\ref{fig4}(c), and Fig.~\ref{fig4}(d), respectively. In Fig.~\ref{fig4}(b) and Fig.~\ref{fig4}(d), it can evidently be seen that the mass loss rates of the covariant quantum BHs are different from those of Schwarzschild BHs for sub-Planckian masses. In addition, the magnitudes of the mass loss rates of the covariant quantum BHs induced by the emission of massless neutrinos are significantly larger than their Schwarzschild counterparts and those for other spin scenarios in both quantum and classical cases in the low-mass regime, whereas those for gravitons are smaller than the Schwarzschild results. However, in Fig.~\ref{fig4}(c), we can find the mass loss rates of covariant quantum BHs are almost the same as their Schwarzschild counterparts. This shows the quantum corrections are almost negligible for the photon contribution to mass loss rate.

It is worth noting that the semi-analytical result in the previous subsection suggests possible remnant formation, whereas the fully numerical results indicate that no remnants are formed. This discrepancy may be from the approximations required to derive the analytical greybody factors. Especially for the covariant quantum BH metric, the required low-frequency approximation is more restrictive than that for the Schwarzschild metric, which can lead to significant errors in the low-mass regime. Therefore, a more reliable way is to calculate the BH evaporation using a fully numerical method.

\section{Conclusions}
\label{sec:conclusions}
Covariant quantum BHs are an important achievement in LQG. Quantum corrections can modify BH spacetime geometry, which may lead to deviations in BH temperature and greybody factors from the classical limits. Our results suggest that the covariant quantum BHs have the same temperature as Schwarzschild BHs, and the greybody factors are also comparable to those for Schwarzschild BHs in the large-mass BH regime. However, in the low-mass BH regime, the greybody factors for the covariant quantum BHs significantly differ from their Schwarzschild counterparts, except for photons.

The massless scalar field and massless neutrino cases were chosen as examples to draw Fig.~\ref{fig1} and Fig.~\ref{fig2}. Compared with Schwarzschild BHs, the greybody factors of the covariant quantum BH for a massless scalar field are smaller than their classical limits in a certain frequency range as shown in Fig.~\ref{fig1}(a). However, for massless neutrino emission, the greybody factors are significantly larger than those for the other cases at certain frequencies in Fig.~\ref{fig1}(a). In particular, when the frequency $\omega=1$, the quantum corrected greybody factor for massless neutrino emission is several orders of magnitude larger than those of the other results in Fig.~\ref{fig1}(a). Based on these modifications of greybody factors, the spectra of the Hawking radiation of the covariant quantum BH with $M=0.01$ differ substantially from those of the Schwarzschild BH in a certain frequency range as shown in Fig.~\ref{fig2}(a). Especially, when the frequency $\omega=1$, the quantum corrected Hawking radiation for massless neutrino emission is stronger by several orders of magnitude than the corresponding results in the classical limit and the massless scalar field case. In contrast, when BH mass $M=0.5$, the greybody factors in the covariant quantum BH almost return to the classical limits, leading to Hawking radiation spectra comparable to those of the Schwarzschild BH. These results indicate the quantum corrections are weak for large-mass BHs, and they are strong for low-mass BHs in the massless scalar field and massless neutrino cases.

Subsequently, we calculated the mass loss rates for massless scalar field, photon, massless neutrino, and graviton cases in Fig.~\ref{fig4}. From this figure, we found that the magnitudes of the mass loss rates of the covariant quantum BHs are significantly smaller than the Schwarzschild results for massless scalar field and gravitons, when BH mass is below $M=0.4$ and $M=0.3$, respectively. This indicates the evaporation of the covariant quantum BHs is weaker than that of Schwarzschild BHs for spin 0 and 2 cases. Moreover, for photons, the mass loss rates of the covariant quantum BHs are almost the same as their Schwarzschild counterparts, and these results indicate that the quantum corrections are weak.

On the other hand, for massless neutrinos, the magnitudes of the mass loss rates of the covariant quantum BHs are distinctly larger than those of Schwarzschild BHs when BH masses are below $M=0.1$, and the magnitudes are significantly larger than those of other spin fields. In addition, the total mass loss rates of covariant quantum BHs are almost consistent with those in the massless neutrino case, as shown in Fig.~\ref{fig4}(b). This indicates the evaporation of the covariant quantum BHs is dominated by massless neutrino emission in the low-mass regime. Although the semi-analytical result shows the covariant quantum BHs may form remnants, as shown in Fig.~\ref{fig3}, the fully numerical results indicate these BHs do not leave remnants. These results suggest it may not be adequate to consider only the massless scalar field case for the evaporation and fate of quantum-corrected BHs, and the contributions from multiple spin fields may need to be included.

For all spin scenarios considered here, the covariant quantum BHs will finally evaporate completely, while the total evaporation is faster than that of Schwarzschild BHs in the low-mass regime. In addition, this spin-dependent behavior, in which the magnitudes of the mass loss rates of the covariant quantum BHs can be larger than, smaller than, or comparable to those of Schwarzschild BHs, may provide a feasible way to test LQG in the future. 

\section*{Acknowledgments}

This work is supported by National Natural Science Foundation of China (NSFC) with Grant No.12275087.

\appendix

\section{Analytical greybody factors}\label{Appendix A}

It is generally difficult to obtain complete analytical greybody factors, so approximations are usually required~\cite{Harmark:2007jy,Belfaqih:2026qgj}. In this Appendix, the low-frequency approximation $(M\omega\ll1)$ is adopted to derive analytical greybody factors for a massless scalar field. To facilitate analytical calculation, the spacetime outside the BH is often divided into three regions.
\begin{itemize}
  \item \textbf{Region I:} The near-horizon region, where $V(r) \ll \omega^{2}$.
  
  \item \textbf{Region II:} The intermediate region, defined by $V(r) \gg \omega^{2}$.
  
  \item \textbf{Region III:} The region far from the horizon, where $V(r) \ll \omega^{2}$.
\end{itemize}
For region I, Eq.~(\ref{eq:SchEq}) has the near-horizon plane wave solution given by Eq.~(\ref{near-horizon}). In the low-frequency approximation, the solution can be expanded as follows (here and hereafter, the subscripts I, II and III denote regions I, II and III, respectively)
\begin{equation}\label{near-horizon2}
    Z_\mathrm{I} \simeq A_{hor}^{\rm in}-iA_{hor}^{\rm in}\omega r_*,
\end{equation}
while the tortoise coordinate
\begin{equation}\label{eq:tort}
    dr_* = \frac{dr}{(1-\frac{2M}{r})(1+\frac{\zeta^2}{r^2}(1-\frac{2M}{r}))},
\end{equation}
can be expanded near the horizon and integrated to give
\begin{equation}\label{r_*}
    r_*\simeq 2M+\frac{\zeta^2}{2M}-\frac{\zeta^4}{96M^3}+2M\ln\left(\frac{r-2M}{2M}\right).
\end{equation}
After substituting Eq.~(\ref{r_*}) into Eq.~(\ref{near-horizon2}), one obtains
\begin{equation}\label{near-horizon3}
\begin{split}
    Z_\mathrm{I}\simeq A_{hor}^{\rm in}\left [1-i\omega \left(2M+\frac{\zeta^2}{2M}-\frac{\zeta^4}{96M^3}+2M\ln\left(\frac{r-2M}{2M}\right)\right)\right ].
\end{split}
\end{equation}
For region II, considering only the leading mode $l=0$, the most general solution of Eq.~(\ref{eq:SchEq}) is as follows
\begin{equation}\label{region 2 eq}
\begin{split}
    Z_{\mathrm{II}}=r\left [A+B\int^r_\infty \frac{dr'}{r'^2(1-\frac{2M}{r'})(1+\frac{\zeta^2}{r'^2}(1-\frac{2M}{r'}))}\right],
    \end{split}
\end{equation}
where $A$ and $B$ are constants. For determining $A$ and $B$, Eq.~(\ref{region 2 eq}) needs to be expanded near the horizon, yielding
\begin{equation}
    Z_{\mathrm{II}}\simeq2AM+\frac{B(40M^2-\zeta^2)\zeta^2}{480M^4}+B\ln\left(\frac{r-2M}{2M}\right).
\end{equation}
To match the wave function of region $\mathrm{I}$, the coefficients $A$ and $B$ are determined as follows

\begin{equation}
    A= A^{in}_{hor}\left[\frac{1}{2M}-i\omega-\frac{i(160M^2-3\zeta^2)\zeta^2\omega}{960M^4}\right],
\end{equation}
and 
\begin{equation}
    B=-2i\omega MA^{in}_{hor}.
\end{equation}
For matching the solution in region III, it is necessary to know the asymptotic behavior of the solution in region II far from the horizon. When $r\gg r_H$ and $r\gg \zeta$, Eq.~(\ref{region 2 eq}) can be rewritten as
\begin{equation}\label{region 2 far}
\begin{split}
   Z_{\mathrm{II}}&\simeq r\left (A+B\int^r_\infty \frac{dr'}{r'^2}\right)
   \\
   &=A^{in}_{hor}\left(\frac{r}{2M}+2iM\omega -ir\omega -\frac{ir\zeta^2 \omega}{6M^2}+\frac{ir\zeta^4\omega}{320M^4} \right).
\end{split}
\end{equation}
For region III, Eq.~(\ref{eq:SchEq}) has the plane wave solution Eq.~(\ref{Zfar}). In the low-frequency approximation, the solution can be expanded as follows
\begin{equation}\label{zIII}
    Z_{\mathrm{III}} \simeq A^{in}_{\infty}+A^{out}_{\infty}-i\omega r_*(A^{in}_{\infty}-A^{out}_{\infty}).
\end{equation}
To match Eq.~(\ref{region 2 far}) with Eq.~(\ref{zIII}), the coefficient $A^{in}_{\infty}$ is solved for as follows
\begin{equation}
    A^{in}_{\infty}=A^{in}_{hor}\left(\frac{1}{2}+\frac{\zeta^2}{12M^2}-\frac{\zeta^4}{640M^4}+\frac{i}{4M\omega}+iM\omega\right).
\end{equation}
Using Eq.~(\ref{greybody factors}), the final analytical greybody factor is given by
\begin{equation}\label{analitical greybody factors2}
\Gamma=\frac{16M^2\omega^2}{M^2\omega^2(2+\frac{\zeta^2}{3M^2}-\frac{\zeta^4}{160M^4})^2+(1+4M^2\omega^2)^2}.
\end{equation}
When $\zeta\rightarrow0$ and $M\omega \ll1$, Eq.~(\ref{analitical greybody factors2}) reduces to the Schwarzschild case as follows~\cite{Harmark:2007jy}
\begin{equation}\label{Sch greybody factors}
    \Gamma \simeq 16M^2\omega^2.
\end{equation}

\bibliographystyle{JHEP}
\bibliography{ref}

\end{document}